\documentclass{article}
\usepackage{arxiv}
\usepackage[utf8]{inputenc}
\usepackage{arxiv}
\usepackage{cite}
\usepackage{nameref,hyperref}
\usepackage{graphicx}
\usepackage[right]{lineno}
\usepackage{float}
\usepackage{microtype}
\DisableLigatures[f]{encoding = *, family = * }

\raggedright
\usepackage{changepage}

\usepackage[aboveskip=1pt,labelfont=bf,labelsep=period,singlelinecheck=off]{caption}

\makeatletter
\renewcommand{\@biblabel}[1]{\quad#1.}
\makeatother

\usepackage{lastpage,fancyhdr,graphicx}
\usepackage{epstopdf}
\fancyheadoffset[L]{1.25in}
\fancyfootoffset[L]{1.25in}

\usepackage{color}
\usepackage[T1]{fontenc}

\definecolor{Gray}{gray}{.25}

\providecommand{\keywords}[1]
{
  \small	
  \textbf{\textit{Keywords---}} #1
}
\usepackage{wrapfig}
\usepackage[pscoord]{eso-pic}
\usepackage[fulladjust]{marginnote}
\reversemarginpar

\renewcommand{\shorttitle}{\textit{arXiv} Template}

\begin{document}
 
\begin{flushleft}
{\Large\textbf{Improving Recycling Accuracy across UK Local Authorities: A Prototype for Citizen Engagement 
\\
A Pre-print}}

\hrulefill

Chloé Greenstreet \textsuperscript{1},
Anastasia Vayona  \textsuperscript{2},
Jane Henriksen-Bulmer \textsuperscript{1*}
\bigskip
\newline
{1} School of Computing and Engineering, Bournemouth University, Poole, UK
\\
{2} Department of Life \& Environmental Sciences, Bournemouth University, Poole, UK
\\
\bigskip
* correseponding@jhenriksenbulmer@bournemouth.ac.uk

\end{flushleft}

\section*{Abstract}
Despite public motivation to recycle, significant barriers hinder effective household recycling in the UK. Decentralised local authority waste management creates citizen confusion and "wishcycling" (disposing of non-recyclable items in recycling bins). The recent Simpler Recycling Policy further complicates this landscape by mandating new identification, sorting, and cleaning requirements that will require citizen guidance to ensure they understand how these will impact their recycling practices.

This mixed methods study (surveys n=50, expert interviews, design activities) used the Value Proposition Canvas to identify citizen pain points: confusion about logos, logistical constraints, and information gaps about local requirements. We then developed an interactive prototype application providing location-specific guidance, visual sorting aids, and material-specific information to address these painpoints.

Focus group evaluation showed the prototype improved recycling accuracy by 60 percent, with marked improvements in packaging assessment. Technology-enabled solutions grounded in user-centred design can measurably improve recycling behaviours and reduce contamination. However, such solutions are most effective when complementing (rather than substituting for) systemic improvements in local authority communication and service design.

\keywords{Household recycling, Citizen behaviour, Wishcycling, Simpler Recycling Policy, Waste contamination, Sustainability, Prototype, User-Centered Design}

\section*{Introduction}
Local government waste management services face mounting pressure to deliver effective recycling systems that meet both environmental targets and citizen expectation \cite{Mollah2025}. Although public awareness and motivation to recycle have grown in recent years, significant barriers within local government service delivery continue to hinder effective recycling practices among UK households \cite{Shooshtarianeetal_2023}. These barriers are compounded by the structural complexity of waste management governance in the United Kingdom (UK, where responsibility for collecting and informing citizens about recycling is delegated to individual local authorities (LAs), each determining the specifics of what and how they recycle materials within their jurisdiction \cite{bashir2022}.

This decentralised approach has resulted in substantial regional variation in recycling services across the UK \cite{dewasiri2025} . Differences in accepted materials, collection frequencies, and preparation requirements between LAs contribute to widespread citizen confusion regarding what can be recycled at the kerbside and when \cite{Oluwadipe2022}. The variation stems from LAs’ distinct Material Recovery Facility (MRF) contracts \cite{bashir2022}, creating a patchwork system that challenges both service providers and users. 

Beyond kerbside collection, LAs face additional service delivery challenges in providing access to specialised recycling facilities. Limited access to recycling drop-off points and specialised programs, make it difficult for citizens to dispose of items that are not commonly recyclable at the kerbside, such as batteries, electronic waste, and disposable vapes \cite{rahman2025}. The accessibility of these collection points varies, depending on legislation and policies that mandate their provision. For example, the Waste Batteries and Accumulators Guidance and initiatives like the Flexible Plastic Pack \cite{WRAP2025}, enforce the provision of collection points for batteries and flexible plastics at major supermarkets across the UK. However, beyond these mandated facilities, additional recycling points hosted by individual institutions within local communities are not regulated, resulting in uneven distribution of services \cite{Oluwadipe2022}. For example, recycling a child’s lunchbox requires visits to multiple collection points, demonstrating how service accessibility challenges may exacerbate citizens’ willingness to engage.

The recent Simpler Recycling Policy, which comes into force in March 2026, has placed additional demands on both services and householders, mandating new identification, sorting, and cleaning requirements \cite{DEFRA2024}. Without adequate guidance, the policy risks exacerbating existing confusion and widening the gap between policy intentions and citizen compliance. 

These service delivery challenges manifest in problematic citizen behaviours that undermine recycling system effectiveness. “Wishcycling”, the practice of placing questionable or non-recyclable items into recycling bins in the hope they will be recycled \cite{vayona2024}, exemplifies how inadequate LA information provision and service accessibility contribute to recycling stream contamination. This avoidable behaviour has contributed to an estimated 2/3 of packaging waste being diverted to landfill \cite{McGrath_BBC2018}. Understanding and addressing the service delivery gaps that drive such behaviours is essential for improving LA recycling outcomes and achieving waste management targets.

\subsection*{Research Gap and Objectives}
While existing research has documented barriers to household recycling \cite{Oluwadipe2022}, limited attention has been paid to systematically analysing these from LA service delivery perspective or developing practical interventions that LAs can implement. 
This study addresses these gaps by examining recycling service delivery through the Value Proposition Canvas \cite{osterwalder2015}, a framework adapted here to analyse citizen needs and pain points in public service delivery, using this framework to develop an interactive prototype application to help citizens improve their recycling practices. The study is guided by the following research question (RQ): \emph{"How can a smartphone application reduce wishcycling and improve the quality of household recycling within inconsistent local authority systems?"}.

This study focuses on recycling behaviours within UK households in relation to LA services, excluding food waste management and construction material recycling, as they involve different regulatory frameworks.

\section*{Background}
\subsection*{Local Government waste management in the UK}
In the UK, the Environmental Protection Act 1990 delegates responsibility for waste collection and disposal to LAs, including provisions for material separation \cite{mohammed2021}. Thus, waste management represents a critical LA service, with substantial implications for environmental sustainability, public health, and LA finances. Further, this decentralised governance structure positions LAs as primary actors in implementing national environmental policy at the community level, requiring them to balance statutory obligations, resource constraints, and citizen expectations. 

In addition, responsibility for implementing the provisions of Climate Targets introduced by the EU mandating 55\% emissions reductions by 2030 \cite{EC2020} have cascaded down to LA service delivery, requiring councils to enhance household recycling participation, despite limited resources and varying local conditions. These obligations, coupled with the 2030 Agenda for Sustainable Development and its 17 Sustainable Development Goals (SDGs) \cite{UNSDG}, adopted by UN member countries, including the UK, in 2017 \cite{DIT2017}, have intensified policy pressure on LAs to improve recycling outcomes. Yet despite heightened public environmental awareness resulting from these initiatives, LAs face persistent challenges in delivering effective recycling services that translate citizen motivation into successful recycling behaviour.

\subsection*{Circular Economy and the Waste Hierarchy}
\label{sec:2b}
LA waste management services operate within the broader framework of the circular economy (CE), an economic model that emphasises keeping resources in use for as long as possible, extracting maximum value during use, and recovering materials at the end of product life \cite{EMF2015}. Unlike the traditional linear “take-make-dispose” model, CE seeks to close resource loops through reduction, reuse, repair, refurbishment, remanufacturing, and recycling. For LAs, transitioning toward CE principles requires not only managing end-of-life waste, but also encouraging citizen behaviours that prioritise waste prevention and material longevity \cite{puvavca2025}.

The waste hierarchy provides the operational framework for CE implementation in LA contexts \cite{awino2024}. Recycling, defined as \emph{“the repurposing in a production process of waste materials for the original purpose or for other purposes”}\cite{Recycling2004}, sits within this hierarchy, which prioritises prevention (reducing consumption) and reuse (extending product life) above recycling as more effective strategies for waste reduction. The widely recognised phrase “reduce, reuse, recycle” was originally intended to reflect this order of importance, acknowledging that preventing waste generation delivers greater environmental benefits than processing waste after creation \cite{mohammed2021}.

However, LA service delivery focuses predominantly on the recycling component of this hierarchy \cite{anshassi2021}. While citizen recycling of plastics, paper, glass, and metals represents the core LA waste management offering, this emphasis on recycling potentially overshadows higher-priority CE activities \cite{BONGERS2022}. Citizens may perceive recycling as the primary pro-environmental behaviour, neglecting reduction and reuse practices  that would reduce waste streams and resource consumption more significantly. For LAs seeking to advance CE transitions, this presents a dual challenge: improving recycling service effectiveness, while elevating citizen awareness of waste prevention and reuse practices positioned higher in the hierarchy \cite{konstantinidou2024}.  

\subsection*{Service delivery fragmentation}
A defining characteristic of UK waste management is the substantial variation in service delivery models across LAs. Each LA determines what materials are accepted for collection, preparation requirements, collection frequencies, and sorting expectations for citizens (Fontaine et al. 2024). This variation stems from LAs distinct MRF contracts, that dictate which materials can be economically processed in each area \cite{bashir2022}. While this flexibility allows LAs to adapt services to local infrastructure and market conditions, it creates significant challenges for citizens attempting to understand and comply with recycling requirements.

The complexity of inter-council variation extends beyond material types to encompass area-specific guidelines and preparation requirements. Different LAs impose varying expectations on citizens, with some LAs requiring householders to rinse containers or remove labels, while others stipulate no such preparation \cite{lee_etal_2022}.  Accepted materials also differ substantially across LAs. For instance, some LAs accept plastic film while others prohibit it. A RECOUP survey revealed substantial citizen confusion regarding which plastics their LA accepts \cite{McBeth_2023_RECOUP}, suggesting that service fragmentation creates information barriers that undermine compliance. For householders who relocate between LA jurisdictions, previously established recycling habits may no longer align with new local requirements, leading to unintentional non-compliance and contamination.

\subsection*{Service delivery gap}
The interaction between LA service design and citizen behaviour reveals critical gaps in current waste management systems. Despite more than five decades of recycling initiatives, many UK citizens struggle to understand recycling requirements, resulting in “wishcycling”, placing non-recyclable items into recycling bins hoping they will be processed \cite{piper2022rubbish}. This behaviour, driven by pro-environmental motivation but inadequate information, undermines recycling system effectiveness: only 25.5\% of plastic waste is recycled in England, with approximately 55\% entering residual waste streams for disposal to landfills, Energy from Waste facilities, and Mechanical Biological Treatment plants, while the remaining 19.5\% becomes litter \cite{hahladakis_etal_2018}.  

Contamination represents one of the most persistent operational challenges for LAs \cite{timlett2008}. Before materials can be recycled, householders must remove contaminants such as food residue, liquids, or non-recyclable components. Contamination occurs through improper cleaning (e.g., pizza boxes with grease), storage issues, or weather exposure. Each contaminated item increases processing costs and reduces the market value of recovered materials \cite{price2020}.  From a LA perspective, citizen education and service design that prevents contamination represent critical determinants of system efficiency and cost-effectiveness.

\subsection*{Specialist collection infrastructure}
Beyond kerbside services, LAs face challenges providing accessible collection infrastructure for materials not economically viable for routine collection \cite{henry2006}. The decision regarding which materials are collected kerbside is driven partly by commercial value. As manufacturers commit to using recycled content, demand for recyclable materials increases, enhancing market value and justifying collection costs for common materials like cardboard, glass, and plastic \cite{dhawan2022}. Conversely, less valuable materials (polystyrene, textiles, composite items) require specialist collection points that are difficult to provide widely.

Further, beyond mandated facilities, additional collection points hosted by schools, community centres, or retail outlets operate without regulation, resulting in uneven geographic distribution influenced by socio- economic factors \cite{Oluwadipe2022}. This creates disparities in service accessibility, with some communities having convenient access to recycling infrastructure, while others face substantial barriers. For items like personal care products, textiles, or composite packaging, locating appropriate collection points can be prohibitively time-consuming, diminishing citizen motivation.

Research by WRAP found that over 60\% of UK citizens are unaware of where to recycle flexible plastics despite supermarket initiatives \cite{WRAP2023}, highlighting how lack of awareness compounds accessibility barriers. From a LA service delivery perspective, this suggests that infrastructure provision alone is insufficient; LAs must also invest in information provision and citizen engagement to ensure citizens can effectively utilise available services.

\subsection*{Technology and local authority innovation}
LAs increasingly explore technology-enabled solutions to address service delivery gaps and improve citizen engagement with waste management services \cite{pehlken2024}. Digital tools offer potential to overcome information barriers created by inter-council variation, providing location-specific guidance that adapts to individual LA requirements. Technology can also support behaviour change by reducing cognitive burden, offering timely reminders, and making complex information accessible at the point of decision-making \cite{gibovic2021}. 

Moreover, technology interventions present opportunities to promote higher-priority CE behaviours, 
alongside recycling information, potentially rebalancing citizen focus toward more impactful waste prevention activities \cite{bucker2025}.  However, LA technology adoption faces resource constraints and implementation challenges. Solutions must be cost-effective, scalable across diverse contexts, and accessible to populations with varying digital literacy. Moreover, technology interventions must address underlying behavioural and structural barriers, rather than simply digitising existing inadequate information provision. Understanding citizen needs and pain points, what frustrates them, what information they require, and how they make recycling decisions, is therefore essential for designing effective technology solutions that genuinely improve service outcomes, rather than adding digital complexity to already confusing systems.

The Simpler Recycling Policy, that will take effect from March 2026, brings in new identification, sorting, and cleaning requirements that LAs and citizens must meet \cite{DEFRA2024}, thereby further complicating an already fragmented service delivery. This, we contend, risks exacerbating existing confusion and widening the gap between policy intentions and citizen compliance unless better guidance and advice is provided to citizens in a consistent manner and format. 

\subsection*{Valuebased frameworks for understanding citizen needs}

Systematically understanding citizen needs in public service contexts requires analytical frameworks that can identify pain points, desired outcomes, and decision-making contexts. The Value Proposition Canvas, developed for commercial innovation, offers a structured approach to mapping customer jobs-to-be-done, pains, and gains, then designing solutions that address identified needs \cite{osterwalder2015}. While originally a business tool, its systematic methodology for understanding user needs has potential applicability to public service design, particularly in complex service domains like waste management where citizen confusion and service delivery fragmentation create multiple pain points requiring targeted intervention.

\section*{Materials and Methods}
The methodology for this study adopts a mixed methods approach, combining quantitative and qualitative data. This approach was selected as it allows for wide-ranging understanding (through survey data across 50 participants), and depth of understanding (through qualitative interviews, focus groups, and design activities with specialists and end-users). For research addressing complex service delivery barriers in public sector contexts, mixed methods design is particularly valuable, as it enables methodological pluralism across data sources and validates findings across different methodological frames \cite{hendren2023}. More specifically:
\begin{itemize}
\item Quantitative components (surveys): Provides breadth by identifying patterns in recycling behaviours and perceptions across a sample population. While the sample size of 50 participants is modest, this reflects realistic constraints in public sector research and is defensible within mixed methods approaches where qualitative depth compensates for quantitative breadth \cite{edwards2021}. 
\item Qualitative components (interviews, focus groups, design activities): Provide depth by exploring the why and how of recycling barriers. Expert interviews clarify systemic and regulatory factors; focus group evaluation assesses whether designed solutions address identified barriers; think-aloud protocols and card sorting reveal mental models and usability considerations.
\item Design research component (wireframes, prototyping, iterative testing): Translates findings into practical solutions while validating whether design interventions effectively address identified pain points.
\end {itemize}

This integration allows the research to answer not only “what barriers exist” but also “how do these barriers manifest in citizen behaviour” and “can technology-enabled solutions effectively address them”?

\subsection*{Ethics}
Prior to data collection, ethics approval obtained from the University’s ethics committee (Ref No: 61956). Participants provided written informed consent prior to participation, understanding the study’s purpose, data collection methods, confidentiality protections, and right to withdraw at any time. 

\subsection*{User Centered Design Process}

\begin{figure}[ht] 
\includegraphics[width=\textwidth]{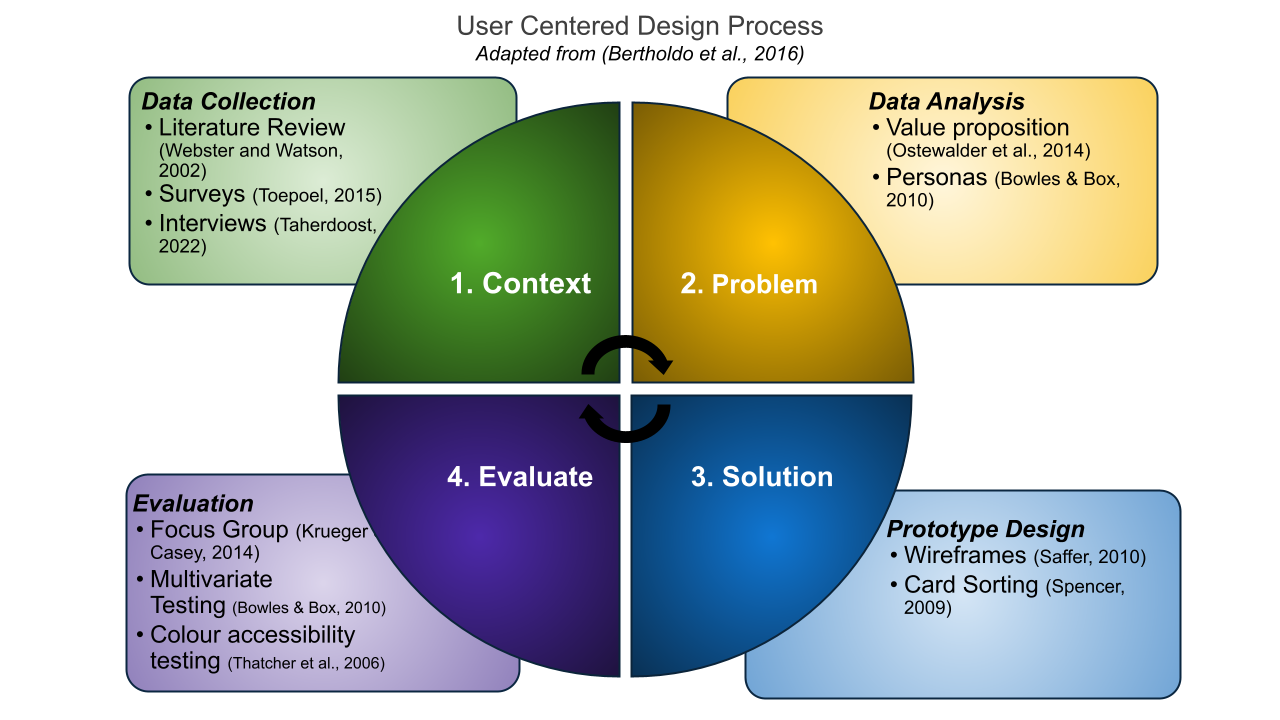}
\caption{\color{Gray} \textbf{User Centered Design Process}.}
\label{fig1} 
\end{figure}

To achieve that, this study employs an Agile User-Centered Design (UCD) methodology \cite{Bertholdo2016}, adapted for public service contexts (Figure \ref{fig1}). 

This iterative, cyclical approach allowed us to systematically identify and address citizen barriers in LA recycling service delivery, while maintaining continuous user involvement throughout the design process. 

\section*{Data Collection \& Analysis}

\subsection*{Surveys}
The study began with an online survey, to examine patterns of household recycling engagement across the UK. Recruitment was facilitated through community-based posters and social media channels, aiming to achieve a diverse respondent base. The survey comprised two sections: Section A focused on perceptions of recycling, incorporating targeted items to identify “wishcycling” behaviours, 
while Section B captured self-reported recycling practices. To situate these behaviours within broader service contexts, the survey additionally included questions on respondents’ views of local recycling policies and waste management provision.

\subsubsection*{Survey Results}
\begin{figure}[ht] 
\includegraphics[width=\textwidth]{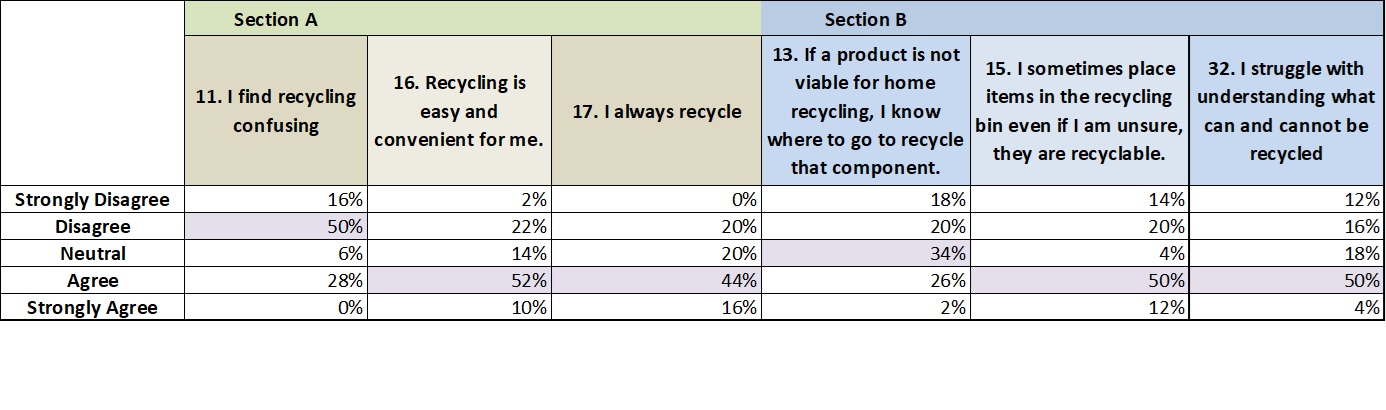}
\caption{\color{Gray} \textbf{Survey results attitude vs practice}.}
\label{fig2} 
\end{figure}

The survey yielded 50 responses, analysed to identify patterns in recycling behaviours and perceptions. Findings revealed a gap between attitudes and practices. While most respondents described recycling as easy and convenient, half reported uncertainty about what materials could be recycled, often engaging in ‘wishcycling’ (Figure \ref{fig2}). This empirical evidence reinforces prior research on persistent knowledge deficits despite positive attitudes \cite{piper2022rubbish}\cite{vayona2024}.

\begin{figure}[ht] 
\includegraphics[width=\textwidth]{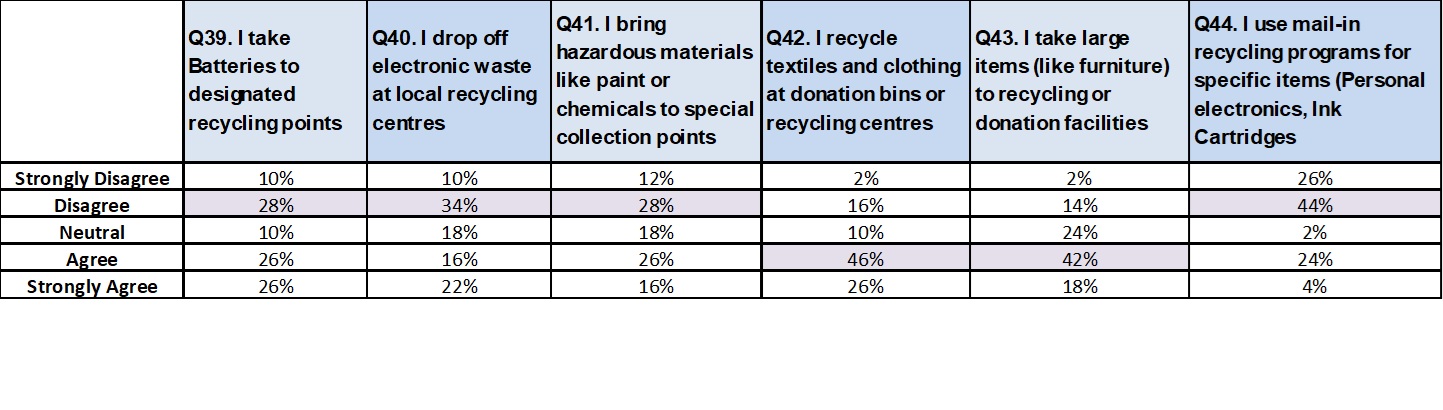}
\caption{\color{Gray} \textbf{Survey results specialist recycling point engagement}.}
\label{fig3} 
\end{figure}

Responses also highlighted the role of accessibility in shaping recycling behaviour. Respondents reported higher engagement with familiar collection points, particularly for clothing and large furniture items, suggesting convenience as a key driver (Figure \ref{fig3}). 
Conversely, mail-in recycling schemes such as First Mile \cite{FirstMile} attracted negative perceptions, suggesting limited awareness or perceived inconvenience.

For high-risk items, such as batteries and electronic waste, responses indicated inconsistent commitment, with standard deviations of 1.37 and 1.33 respectively. Despite recognising hazardous materials as a recycling priority, many respondents lacked clarity on what could be recycled and where, underscoring the need for clearer guidance and improved access to collection facilities.

\subsection*{Interviews}
To complement survey findings, four semi-structured interviews were conducted with industry experts in waste recycling and sustainability. Expert participants represented diverse roles, including LA waste management officers, sustainability specialists, and environmental consultants, with experience ranging from one to six years. Interviews were conducted via Zoom, lasting approximately one hour. Sessions were recorded and transcribed for analysis.

\subsubsection*{Interview Results}
Interview analysis revealed recurring themes, notably confusion caused by recycling logos, identified as a significant barrier to effective recycling by one of the experts who stated: \emph{“The main barrier is the logos. People do not understand them, and they shouldn't have to. You shouldn't need to be an expert to determine if something is recyclable.”} (E2). 

E2 further noted that weak enforcement of logo regulation within the manufacturing industry, contributes to misleading packaging. Some manufacturers intentionally use recycling symbols as a marketing strategy, resulting in contaminated recycling streams driven by both greenwashing and citizens’ misinterpretation of logo meaning \cite{vayona2024}.

Another theme concerned physical and logistical constraints. For example, storage limitations were frequently cited as barriers to proper recycling, with three of the four experts reporting that “wishcycling” often occurs due to limited space in general waste bins. This behaviour was linked to collection schedules and bin capacity: \emph{“Rubbish and recycling are collected every 14 days. Many people, lacking other space, use the recycling bins because they have no other option [creating] a significant barrier to establishing an effective recycling system”} (E2).

\subsection*{Value Proposition}
\begin{figure}[H] 
\includegraphics[width=\textwidth]{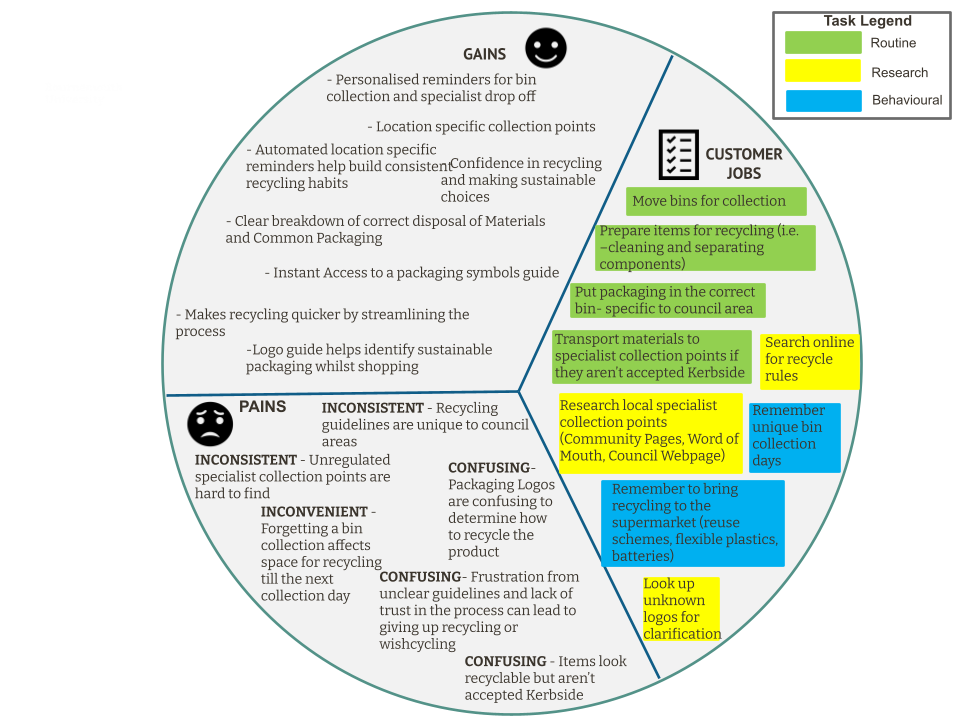}
\caption{\color{Gray} \textbf{Customer Profile (CP)}.}
\label{fig4} 
\end{figure}

\begin{figure}[H] 
\includegraphics[width=\textwidth]{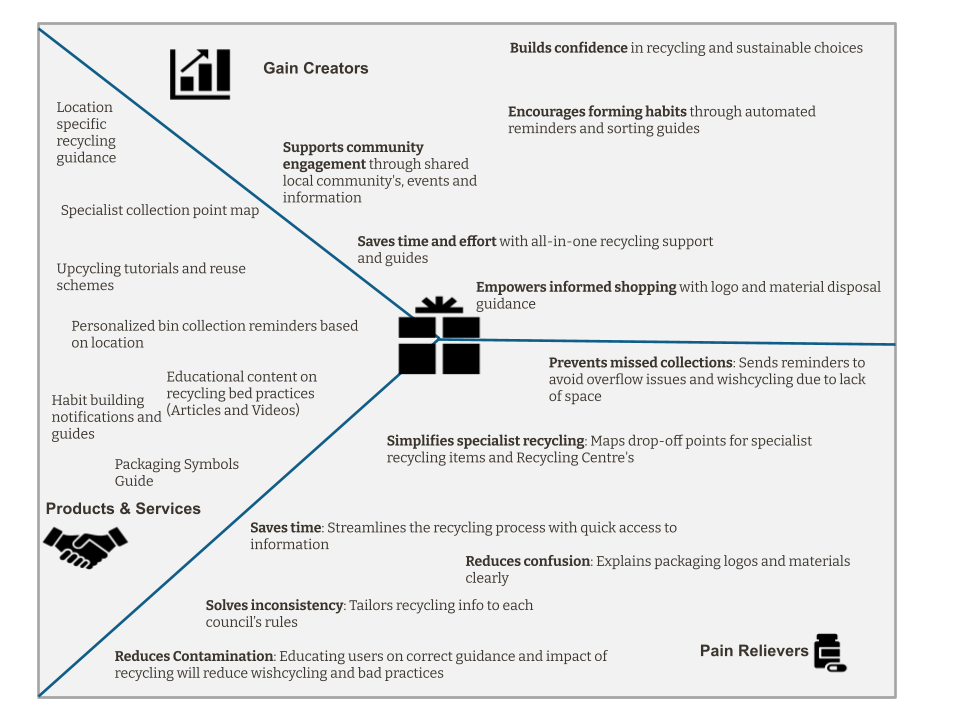}
\caption{\color{Gray} \textbf{Value Map (VM)}.}
\label{fig5} 
\end{figure}

To address citizen pain points identified, a Value Proposition was applied to first map customer needs (Customer Profile (CP), Figure \ref{fig4}) and then translate these into functional requirements for the proposed solution (Value Map (VM), Figure \ref{fig5}) \cite{osterwalder2015}.

The CP illustrates how routine tasks (“customer jobs”) embedded in daily life shape recycling behaviours. These habit-driven tasks generate pain points when disrupted, for example, forgetting scheduled bin collections causes inconvenience. More complex challenges arise when citizens relocate between boroughs and encounter unfamiliar recycling guidelines. 

In such cases, previously established habits can lead to confusion and misinformation regarding proper disposal practices. These behaviours are not indicative of negligence, rather they reflect systemic inconsistencies in recycling rules across the UK, placing the burden of adaptation on citizens, contributing to unintentional errors and diminished confidence in the recycling process.

The VM translates these pain points and desired gains into functional solution components, outlining desired features such as location-specific recycling guidance and educational content, designed to reduce confusion, build confidence, and support habit formation.

\subsection*{Personas}
Building on the CP, three user personas were derived through qualitative analysis of interview and survey data, ensuring the prototype responds to real-world behaviours and barriers. 

\begin{figure}[H] 
\includegraphics[width=.5\textwidth]{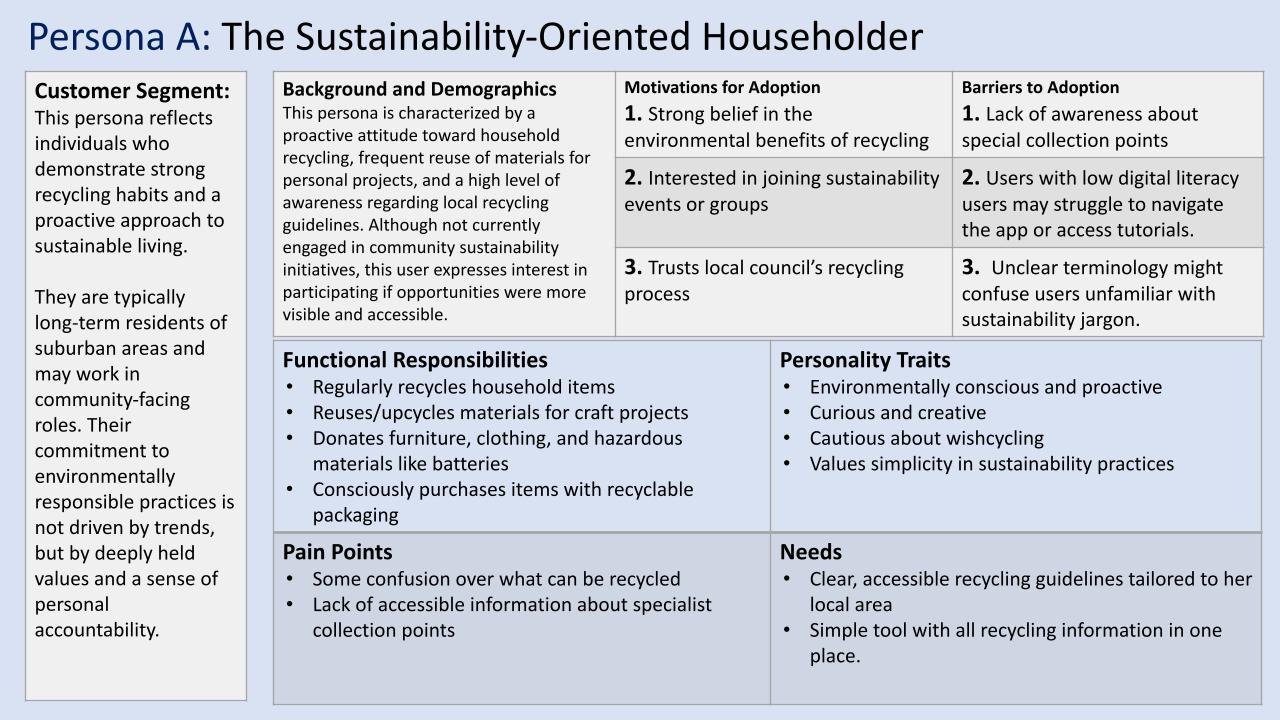}
\caption{\color{Gray} \textbf{Persona A: the sustainability oriented Householder}.}
\label{fig5a} 
\end{figure}

\begin{figure}[H] 
\includegraphics[width=.5\textwidth]{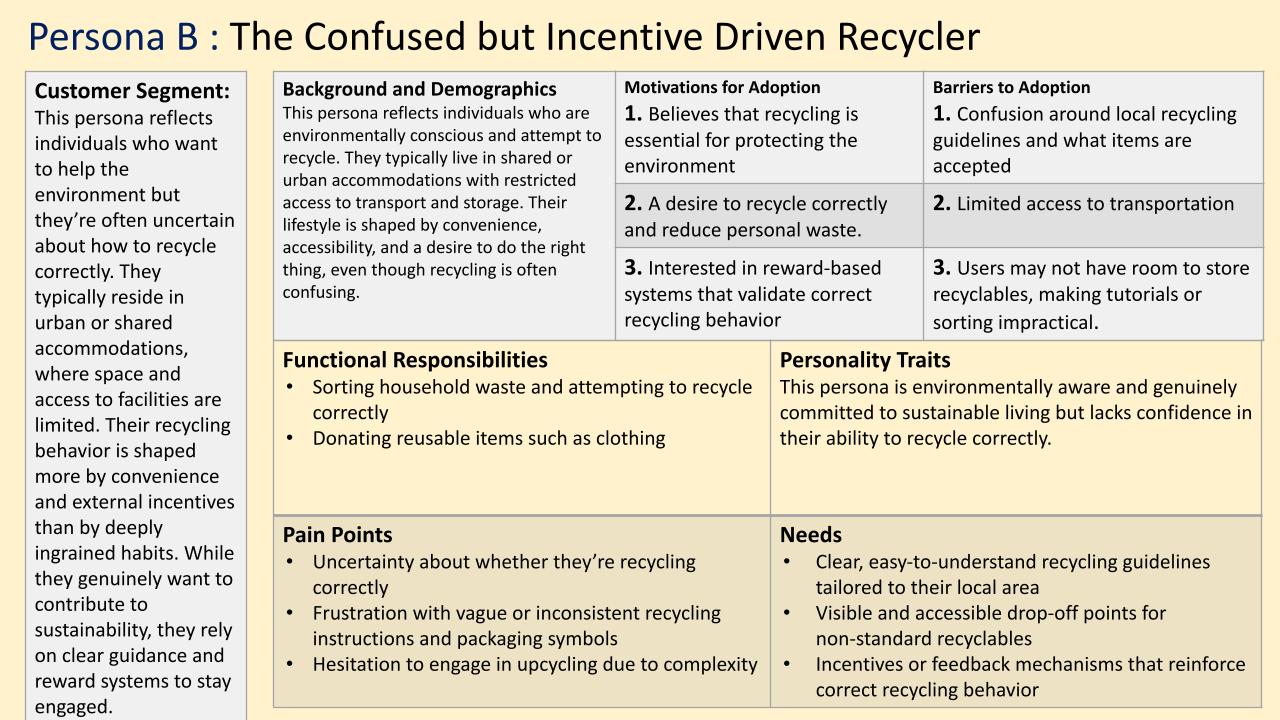}
\caption{\color{Gray} \textbf{Persona B: the confused but Incentive driven Recycler}.}
\label{fig5b} 
\end{figure}

\begin{figure}[H] 
\includegraphics[width=.5\textwidth]{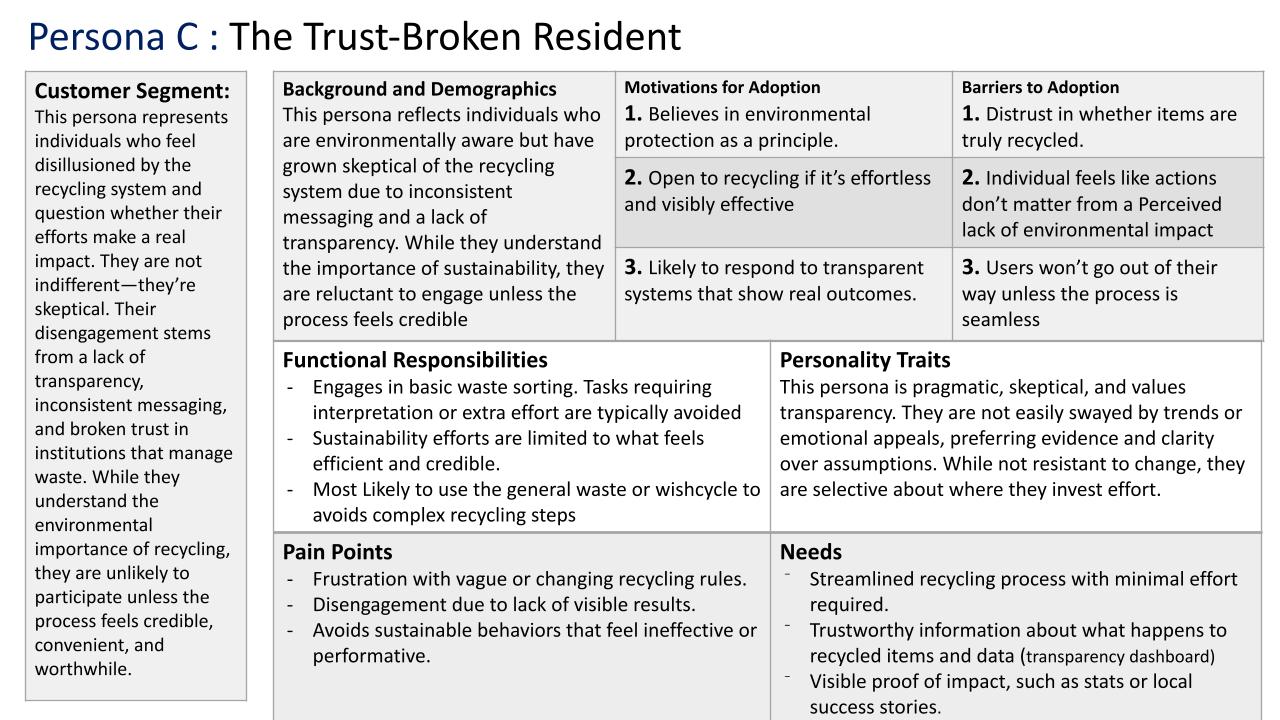}
\caption{\color{Gray} \textbf{Persona C: the trust broken citizen}.}
\label{fig5c} 
\end{figure}

Participants were grouped according to recurring behavioural patterns and stereotypes identified in the interviews, enabling the identification of distinct user types: the sustainability-oriented local (Figure \ref{fig5a}), the confused but incentive-driven recycler (Figure \ref{fig5b}), and the trust-broken citizen (Figure \ref{fig5c}). 

Grounding personas in empirical data rather than assumptions, ensured a final design that reflects specific user needs, motivations, and contextual challenges \cite{bowles2010}.

\subsection*{Wireframes}
\begin{figure}[H] 
\includegraphics[width=.35\textwidth]{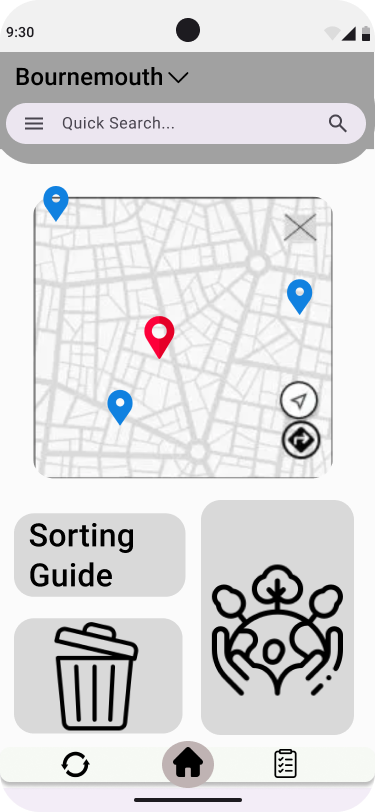}
\caption{\color{Gray} \textbf{Initial Homepage Static Wireframe}.}
\label{fig6} 
\end{figure}

To address behavioural disruptions identified through the pain points (Figure \ref{fig4}), a set of static wireframes was developed to integrate desired features (e.g. Figure \ref{fig6}).

These design elements aim to support citizens in maintaining consistent recycling routines, even when travelling or relocating. By providing timely prompts and contextual guidance, the wireframes seek to reduce missed collections, and prevent improper disposal caused by outdated habits, thereby promoting more confident and informed recycling behaviours across jurisdictions.

The design is able to consolidate recycling rules specific to each LA, addressing a major source of frustration for citizens, the inconsistent and often confusing nature of LA waste guidelines. By streamlining this information into a single accessible platform, correct disposal decisions with greater confidence are facilitated. A further value add is the inclusion of a packaging symbol guide that serves as a reference for recycling and a support tool for sustainable purchasing decisions.

Collectively, these services respond to common problems of misinformation and confusion in household waste management. positioning the system as a practical solution that improves recycling accuracy, while fostering ongoing environmental engagement through locally relevant information and active user participation. 

\section*{Design}
\begin{figure}[H] 
\includegraphics[width=\textwidth]{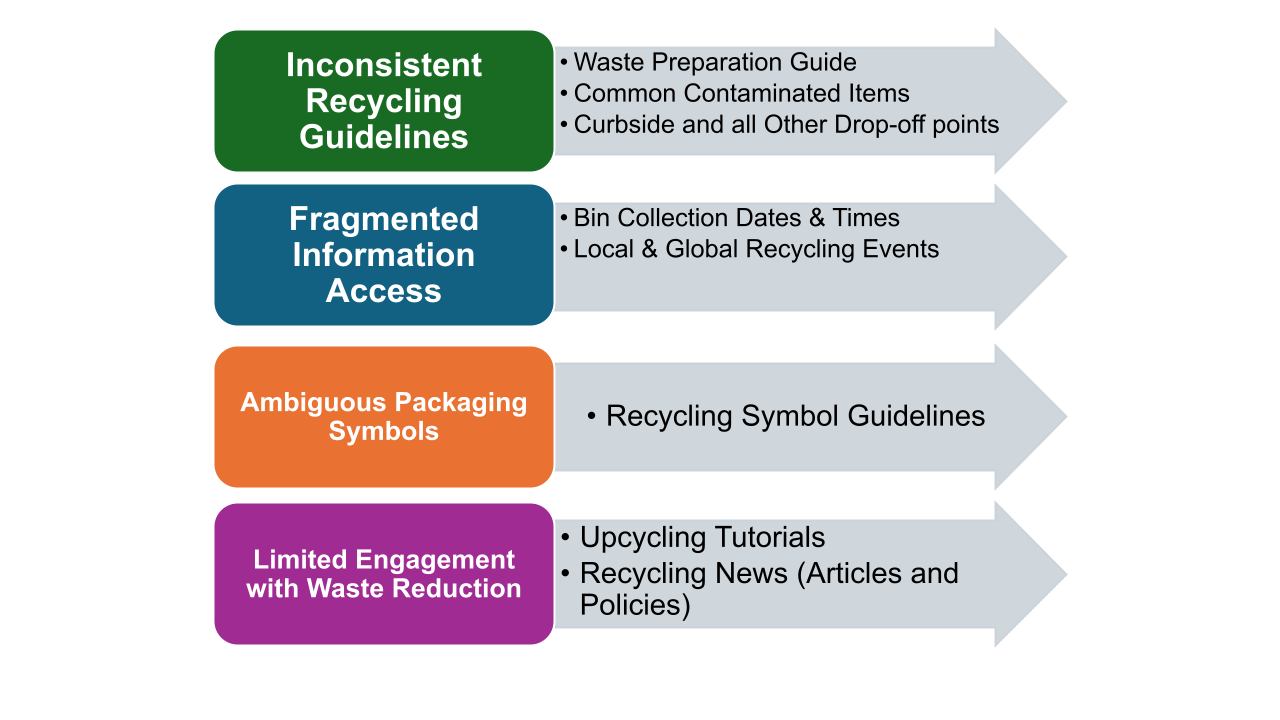}
\caption{\color{Gray} \textbf{Pain Points to Prototype Features}.}
\label{fig7} 
\end{figure}

The next stage was the design process, which involved creating a dynamic, interactive prototype (Figure \ref{fig1}), that would enhance usability and ensure the solution addressed citizen needs effectively. Identified user pain points were systematically translated into functional requirements for the prototype. Challenges such as inconsistent recycling guidelines, fragmented information access, and ambiguity in packaging symbols, were mapped to targeted design features (Figure \ref{fig7})., demonstrating the explicit link between identified barriers and corresponding design interventions, and providing a concise overview of how the prototype seeks to mitigate systemic obstacles to effective recycling.

\subsection*{Open card sorting}
Following the identification of core features (Figure \ref{fig7}), a second round of interviews was conducted via Zoom with three experts (E1, E3, E4), to complete an open card sorting activity \cite{bowles2010}.  Sessions were recorded with consent, and sorting data were automatically captured using the Maze platform.

During these sessions, experts were asked to group, categorise, and label the features identified in the analysis, the aim being to reveal how participants intuitively associate features and uncover underlying mental models. A mental model refers to a user’s internal expectation of system functionality, shaped by prior experiences with similar technologies or tasks. Users evaluate new software by unconsciously comparing it to these models, which can lead to frustration when expectations are unmet \cite{Young_2008}. Verbal explanations and category names were documented to provide additional context before analysing the data to identify common pairings and associations across participants.

Although the sample size was limited, the qualitative nature of the card sorting activity yielded valuable insights into participant expectations \cite{nielsen1994_ThinkAloud}. The strongest correlations revealed consistent groupings, such as the pairing of kerbside pickup items with bin collection schedules which were used to inform dynamic wireframe development, ensuring that design decisions aligned with participant expectations. 

\subsection*{Static wireframe refinement}
The wireframes were refined to ensure the design incorporated accessibility standards \cite{caldwell2008}, usability principles \cite{nielsen1994}, and insights from data collection and card sorting activities. The static wireframes were organised into three primary pages: 

\begin{itemize}
\item Homepage: Serves as central hub, providing quick access to location-specific recycling services, sorting guides, and relevant updates. Its design prioritises immediate engagement and directs users toward key sustainability features. 
\item Upcycling Tutorials: Offers step-by-step guides and materials lists for creative reuse projects. Responding to user interest in sustainable practices while reinforcing the principles of “Reduce, Reuse, Recycle”. 
\item Recycling Material Information: Presents visual sorting aids and material-specific guidance, organised by room or item type. This layout addresses user confusion around packaging and contamination, aiming to reduce “wishcycling” through clear, contextual education.
\end {itemize}

Each screen reflects intuitive feature groupings and navigation pathways identified during the card sorting process. These wireframes provide the basis for further evaluation with participants, enabling systematic improvements to the design through ongoing feedback.

\subsubsection*{Multivariate testing}
In alignment with the methodology (Figure \ref{fig1}), E2 participated in a design variation pilot testing activity to provide feedback on the packaging guideline feature, building on their earlier insights regarding recycling logos. This stage aimed to assess how different page layouts influenced clarity, navigation, and overall usability, ensuring the prototype aligned with user expectations before further refinement. Two wireframe variants were evaluated using a think-aloud approach:

\begin{itemize}
\item Design A a simple, easy to follow linear layout that requires scrolling to access key content.
\item Design B a grouped by recycling logo approach, improving clarity but introducing navigation challenges.
\end {itemize}

\begin{figure}[H] 
\includegraphics[width=0.35 \textwidth]{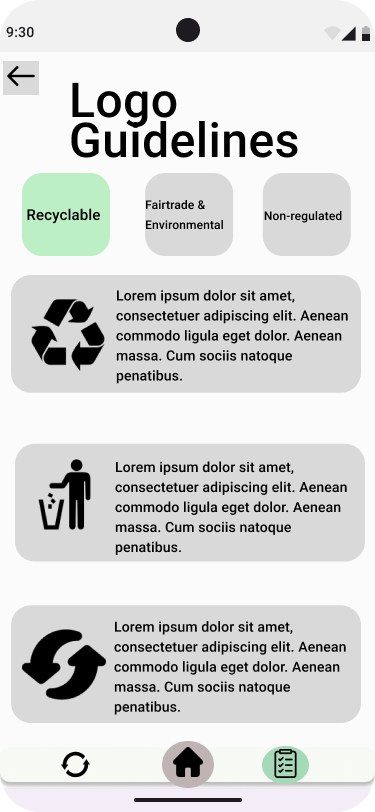}
\caption{\color{Gray} \textbf{Design C guidelines page}.}
\label{fig8} 
\end{figure}

The think-aloud method involves participants verbalising their thoughts while completing tasks, making their reasoning visible for analysis. In this study, the approach was adapted for static wireframes, providing qualitative insights into navigation clarity and content gaps \cite{eccles2017}. Feedback highlighted distinct strengths in each design: simplicity and ease of navigation in Design A, and structured clarity in Design B. Following this feedback, a third layout (Design C) was developed (Figure \ref{fig8}). 

\subsubsection*{ThinkAloud usability testing}
\begin{figure}[H] 
\includegraphics[width=0.35 \textwidth]{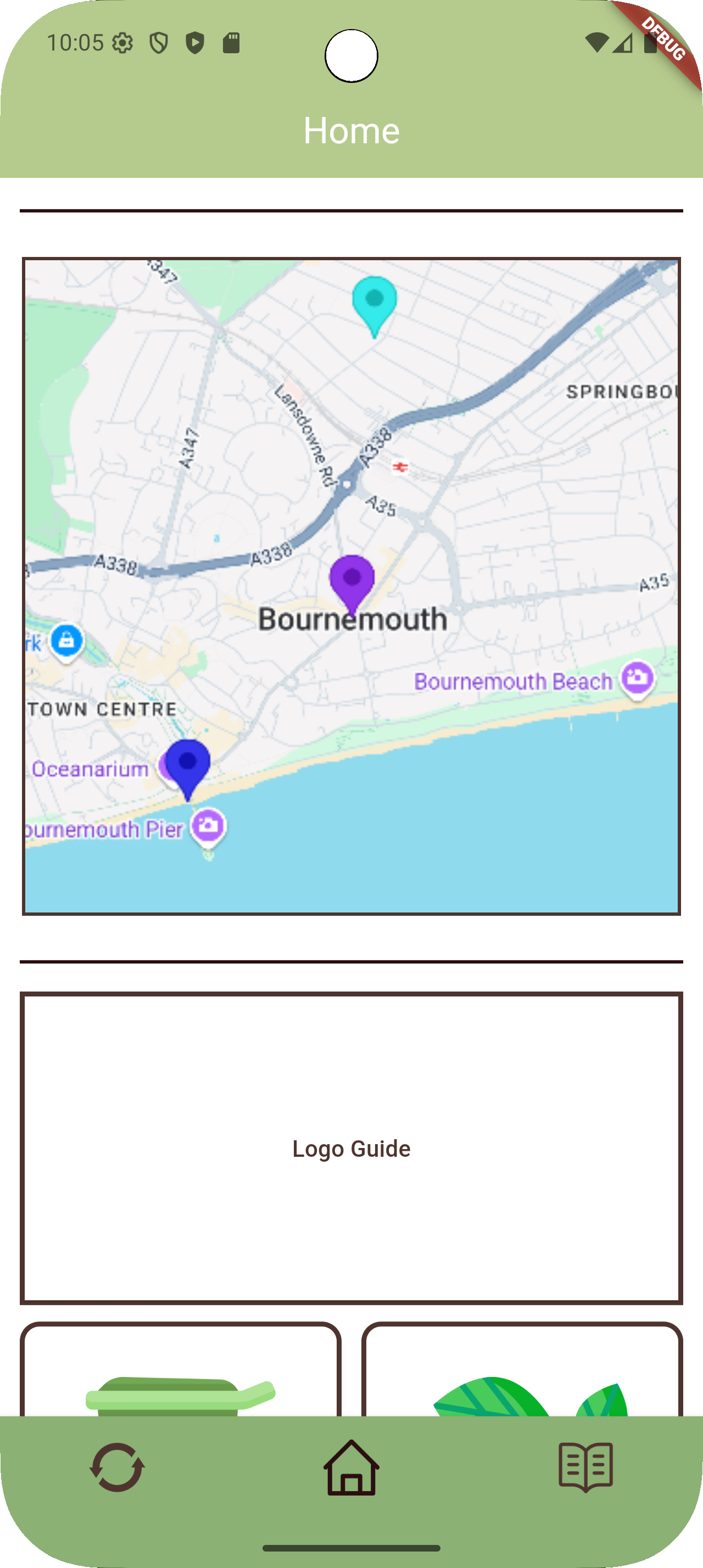}
\caption{\color{Gray} \textbf{Homepage Dynamic Wireframe}.}
\label{fig9} 
\end{figure}

Next, all expert participants took part in think-aloud sessions to evaluate usability and overall user experience. Feedback was generally positive, with participants approving the structure and visual design. Two key recommendations emerged: adding content to the homepage to improve engagement, and integrating the map feature into each guideline page to enhance contextual relevance. These suggestions were incorporated into the updated wireframes (Figure \ref{fig9}).

To consolidate these changes, a navigation diagram was created to illustrate page relationships and linking pathways. The updated wireframes now function as dynamic prototype wireframes, meaning they include interactive elements and navigational flows that simulate real user journeys. 

\section*{Focus Group Evaluation}
\subsection*{Scope of Evaluation}

For the evaluation (Figure \ref{fig1}, stage four), a focus group was conducted, to assess how effectively the prototype addressed the user needs identified. The evaluation examined the dynamic prototype wireframes, to gather qualitative feedback on usability, information architecture, overall user experience and explore whether interaction with the prototype improved participants’ recycling knowledge and accuracy, when compared to baseline performance. Insights from this process informed refinements for the next iteration of the agile cycle (Figure \ref{fig1}).

\subsection*{Focus Group Structure}
The focus group comprised five participants, consistent with Krueger and Casey’s recommendation for optimal group size in qualitative evaluation\cite{krueger2014}. Participants represented Persona A and Persona B (Figures \ref{fig5a} and \ref{fig5b}), which correspond to user profiles with higher engagement levels. No participants aligned with Persona C (low engagement), reflecting the self-selection bias inherent in recruitment through campus channels, and the broader challenge of involving lower-engagement populations in research activities. This absence constitutes a limitation of the study, as it constrained the diversity of perspectives within the evaluation.

The session was structured as a single two-hour event comprising two phases: a control phase and an interactive phase involving the prototype wireframes, enabling pre-post comparison, allowing assessment of whether engagement with the prototype improved participants’ recycling knowledge and accuracy.

\subsection*{Focus Group Structure}
\begin{figure}[H] 
\includegraphics[width= \textwidth]{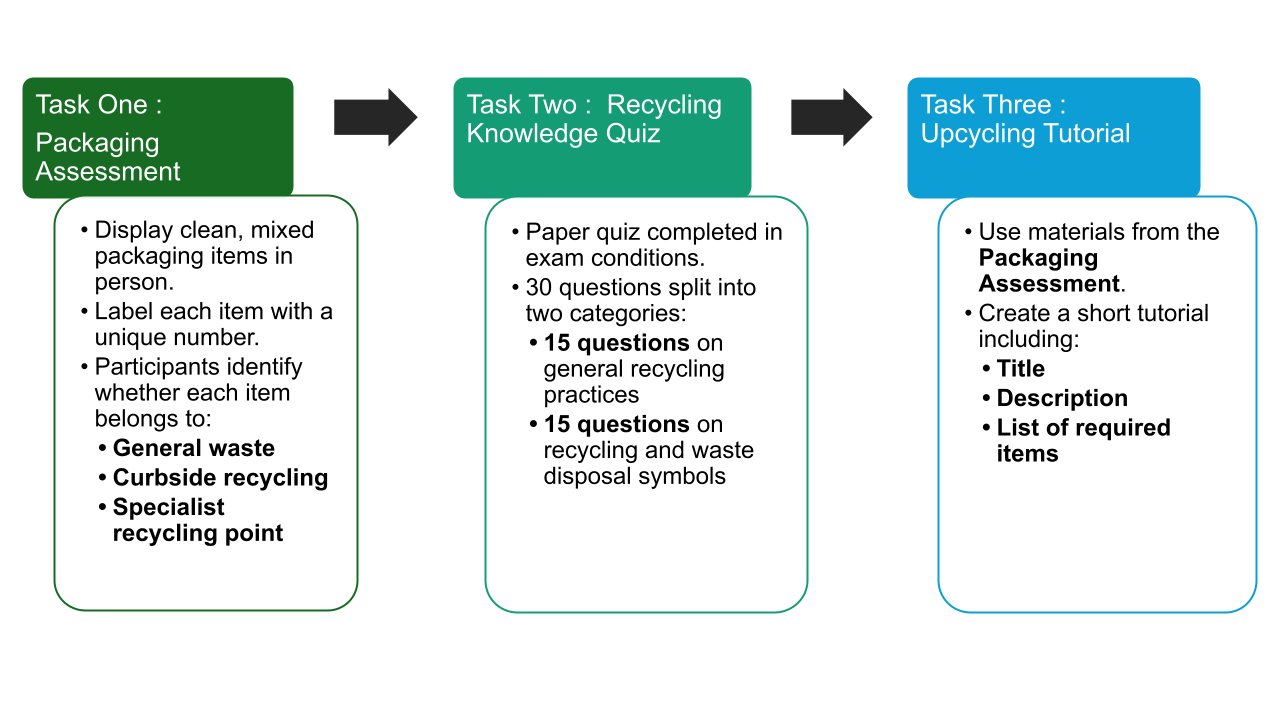}
\caption{\color{Gray} \textbf{Focus Group Tasks}.}
\label{fig10} 
\end{figure}

The focus group evaluation included the same three core activities in both phases, designed to assess participants’ recycling knowledge without and with the aid of the prototype (Figure \ref{fig10}). Feedback was collected throughout both phases to capture participants’ perceptions of usability and effectiveness, informing refinements for the next iteration of development (Figure \ref{fig1}).

\subsubsection*{Control Phase}
In the control phase, participants completed all three tasks without access to the prototype (Figure \ref{fig10}), providing a baseline measure of their existing recycling knowledge, decision-making accuracy, and creative task performance. 

\subsubsection*{Interactive Phase}
The interactive phase replicated the tasks (Figure \ref{fig10}) after participants engaged with the prototype, to determine whether interaction with the prototype improved recycling accuracy, reduced “wishcycling”, and prompted users to reuse their waste through upcycling.

\subsection*{Focus Group Results}
The results showed a clear improvement in accuracy across all activities between the phases. Task One showed the largest increase, with 60\% improvement in answers when having the aid of the prototype. Task Two showed an 18.18\% increase, while Task Three demonstrated a 25\% improvement in correct answers. 
These findings suggest that engagement with the prototype had a measurable positive effect on participants’ recycling knowledge and task performance, across all activities conducted during the focus group.

\subsection*{Participant Feedback}
Feedback from participants corresponding to Persona B indicated that the home page appeared crowded, contrasting with the positive assessments recorded in the experts’ feedback. This divergence likely reflects the experts’ alignment with Persona A, given their expertise in sustainability and recycling. Participants resembling Persona B reported feeling overwhelmed and recommended prioritising essential features such as the map, logo guide, and bin collection schedule on the main screen.

One participant proposed relocating supplementary features to the upcycling tutorial page reasoning that users aligned with Persona A would be more inclined to engage with this content. This recommendation is consistent with findings from the initial card-sorting tests and is expected to enhance overall usability in the next design iteration (Figure \ref{fig1}).

\section*{Discussion}
This research asked; “How can a smartphone application reduce wishcycling and improve the quality of household recycling within inconsistent local authority systems?” How can a smartphone application reduce wishcycling and improve the quality of household recycling within inconsistent local authority systems?

\begin{figure}[H] 
\includegraphics[width= \textwidth]{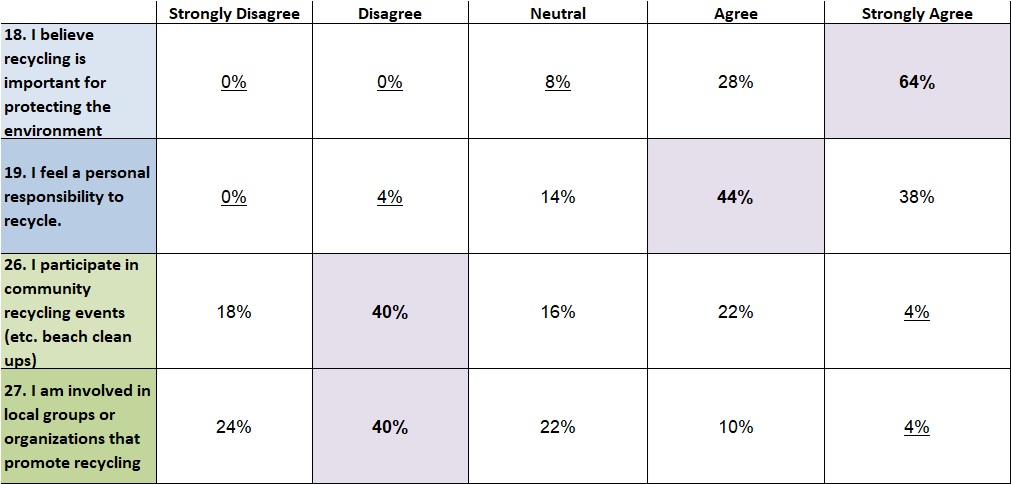}
\caption{\color{Gray} \textbf{Survey Results Recycling Participation Heat Map}.}
\label{fig11} 
\end{figure}

Our findings show that a recycling app can significantly improve the quality of recycling practices and behaviours. These findings demonstrates that a recycling guidance app can measurably reduce wishcycling and improve the quality of household recycling, even within inconsistent local authority systems. Four key findings emerged:

First, survey and interview data revealed a fundamental disconnect between citizen recycling attitudes and behaviours. While 64\% of survey respondents strongly agreed that recycling was important, only 4\% strongly agreed that they participated in local recycling initiatives (Figure \ref{fig11}). This discrepancy suggests there is a disconnect between belief and action by citizens and potentially rooted in a lack of awareness or limited opportunities for engagement communities.

\renewcommand{\thefigure}{2}

This attitude-behaviour gap is not driven by lack of environmental concern, but by systemic barriers in service delivery. Expert interviews identified three recurring obstacles: (1) confusion caused by inconsistent recycling symbols and logos across products, (2) physical and logistical constraints (limited bin capacity, infrequent collection schedules), and (3) information gaps regarding what materials can be recycled locally. These findings align with previous research on recycling barriers in the UK \cite{Oluwadipe2022}, extending understanding by linking barriers directly to local government service delivery Fragmentation.

Second, the focus group evaluation demonstrated that a technology-enabled solution can measurably improve recycling accuracy. Participants showed 60\% improvement on packaging assessment tasks (from 5 to 8 correct answers) after engaging with the prototype, suggesting addressing information barriers through digital means can have direct behavioural impact. Notably, the upcycling tutorial task showed 49.72\% faster completion time with the app, indicating that digital support is particularly effective when users are not juggling multiple reference materials, a practical insight for LA service design.

Third, persona development revealed distinct citizen engagement profiles. While Personas A (high engagement) and B (moderate engagement) participated in the focus group, the absence of Persona C (low engagement) participants highlights a critical service delivery challenge: technology-enabled solutions may primarily benefit already-motivated citizens, while potentially excluding those least likely to engage. This has significant implications for equity and universal service design in LAs.

Fourth, open card sorting showed that citizens intuitively associate features differently than designers expect. For example, participants consistently paired collection schedule reminders with material-specific guidelines, expecting location-based, context-sensitive information. This finding reinforces that user-centered design is essential; generic information provision is insufficient.

\subsection*{Implications for local authorities}

Service fragmentation is one of the structural challenges facing UK LAs, with each LA independently determining which materials are accepted for collection and what preparation is required from citizens, creating a patchwork system of recycling requirements across jurisdictions \cite{harris2024}. This research demonstrates that technology can partially bridge this fragmentation by centralising LA-specific information in a single, accessible platform. Rather than individual LA’s developing separate apps, LAs should consider adopting standardised data formats and contributing to shared, interoperable platforms. 

A consortium approach where LAs collectively contribute standardised data to a centralised platform is more cost-effective than individual solutions, reducing implementation barriers for technology developers, while increasing citizen utility. Critically, LAs should view digital solutions as complementing, rather than replacing, established communication channels. The prototype tested was most effective when integrated with existing LA’s communication infrastructure, suggesting that technology enhances, but does not eliminate the need for effective LA communication strategies. 

Contamination remains one of the most persistent operational challenges for LAs. This research identified specific behavioural drivers of contamination, confusion about which materials can be recycled, uncertainty about material condition requirements, and lack of awareness about where to recycle specialty items. The 60\% improvement in packaging assessment accuracy demonstrates that targeted information provision can directly address these behavioural drivers. For LAs, this has clear cost implications. Processing contaminated materials is expensive, and markets discount contaminated recyclables, reducing the financial return from recycling programs. Expert interviews revealed that “wishcycling” results not from citizens’ ignorance or apathy, but from logistical constraints. As one local expert noted,  \emph{“many people, lacking other space, use the recycling bins because they have no other option”} (E3). Digital tools providing material-specific guidance about washing requirements, acceptable material conditions, and what constitutes contamination at the point of decision, are more effective than traditional marketing materials. The prototype demonstrated that visual aids showing acceptable material conditions are particularly effective. However, technology solutions must be carefully designed to avoid reinforcing existing engagement biases. A critical finding was that technology solutions may primarily benefit motivated citizens, while potentially excluding those least likely to engage. For LAs, where equitable service access is a fundamental responsibility, this limitation is significant. Digital solutions must therefore complement, rather than replace, multiple access methods. Not all citizens have smartphones, digital literacy, or time to engage with mobile applications. LAs should therefore maintain phone support, in-person council offices, and printed materials as complementary access channels. Future research should explicitly include low- engagement populations to ensure digital solutions genuinely serve rather than inadvertently exclude vulnerable citizens.

The 60\%-point gap between environmental values and actual participation in recycling emphasises that citizens perceive recycling as feasible, but waste prevention and reuse as burdensome. Expert interviews reinforced this observation. E1 noted that \emph{“many believe that recycling is better than reusing. It’s already difficult to communicate the importance of recycling, let alone the actions needed for waste prevention”}. Survey respondents disagreed that recycling takes too much time, but agreed that upcycling takes too much time and effort. This suggests that citizens are willing to recycle when barriers are reduced, but lack awareness of or motivation for higher-priority waste prevention activities. 

For LAs, advancing CE transitions should be a priority, as recycling alone cannot achieve waste reduction targets. Digital platforms could integrate waste prevention and reuse information as equally prominent features, not as secondary additions. The prototype intentionally included an “upcycling tutorials” feature, specifically to elevate these higher-priority behaviours and help rebalance public focus toward more impactful waste hierarchy activities, aligning with broader CE goals that extend beyond recycling.

\subsection*{Design Insights for Service Delivery}
The iterative design process produced several actionable insights for LAs, for example, LAs should reorganise information around citizen decision-making moments and specific situations. Our finding showed that Citizens consistently grouped features expecting responses to questions like “I have a battery to recycle, what do I do?", rather than organising information by administrative categories like “hazardous waste”. This mental model, we suggest, could inform the design of both digital platforms and printed materials. 

Visual communication is also critical. Text-based guidelines were significantly less effective than visual examples. Our findings showed that marked improvement could be achieved when visual guidance was provided. Participants consistently requested images showing acceptable material conditions, the difference between clean and contaminated items, and rinsed versus unrinsed containers. Thus, LAs should invest in photography and visual design showing acceptable versus unacceptable material conditions and types with common examples. Finally, localisation is essential. The prototype’s location aware features were consistently praised by participants. Knowing that recycling rules vary by LA, citizens valued having LA-specific information automatically displayed based on their postcode.

\subsection*{Research Limitations}
This research has several important limitations that LAs should consider when implementing findings. The focus group included five participants, skewing toward higher engagement levels, and results cannot be generalised to all UK citizens. However, within mixed methods research, qualitative depth from smaller, purposefully targeted samples are methodologically defensible \cite{krueger2014}. Results demonstrate proof-of-concept that digital solutions can improve recycling accuracy among engaged citizens, and LAs should pilot solutions with their own populations, recognising that improvements may vary by citizen demographic.

Additionally, the focus group measured immediate behaviour change within a single session, but sustained behaviour change requires longitudinal follow-up. LAs should implement built-in analytics to track whether digital tools produce lasting improvements or whether effects fade over time. The absence of low-engagement participants represents a significant limitation. For LAs implementing equitable public services, this gap is critical. LAs should explicitly engage low-engagement populations in service design, recognising that hard-to-reach populations are often those who most need improved service access.

\section*{Conclusion}
This research demonstrates that a smartphone application can measurably reduce wishcycling and improve the quality of household recycling within inconsistent local authority systems. For LAs such solutions offer practical pathways to improve service effectiveness, while managing operational costs associated with contamination. However, technology alone is insufficient. Systemic barriers (service fragmentation, information inconsistency, and collection infrastructure gaps) require policy-level coordination across LAs that extends beyond what individual LA’s can achieve. Technology-enabled solutions are most effective when complementing, rather than substituting, systemic improvements \cite{chiodo2025}.

The most significant contribution of this research is its identification of specific, addressable barriers, confusion about recycling logos, logistical constraints limiting engagement, information gaps about local requirements, and awareness deficits regarding higher-priority waste prevention activities. We have demonstrated that these can be addressed through improved communication, user-centred service design, and strategic technology adoption. The path forward involves strategic, evidence-based improvements to how LAs communicate recycling requirements and support citizen behaviour change. The most impactful outcome would be national guidance standardising recycling symbols, material definitions, and preparation requirements. Until national harmonisation occurs, technology-enabled solutions offer practical, interim pathways to improve service effectiveness and citizen recycling accuracy while supporting broader CE transitions.

\renewcommand{\thefigure}{2}
\setcounter{figure}{0}
\renewcommand{\thefigure}{S\arabic{figure}}

\nolinenumbers

\bibliography{references}

@techreport{WRAP2025,
	author = {{WRAP}},
	howpublished = {[online]},
	institution = {The Waste and Resources Action Programme (WRAP)},
	lastchecked = {Jan 2026},
	month = {Nov},
	title = {The UK Plastics Pact Progress Report 2024/25},
	url = {https://www.wrap.ngo/resources/report/uk-plastics-pact-progress-report-202425},
	year = {2025}}

@article{harris2024,
	author = {Harris, Irina and Bermudez Bermejo, Diego Enrique and Crowther, Thomas and McDonald, James},
	doi = {https://doi.org/10.3390/logistics8040118},
	journal = {Logistics},
	number = {4},
	pages = {118},
	publisher = {MDPI},
	title = {Factors affecting truck payload in recycling operations: Towards sustainable solutions},
	volume = {8},
	year = {2024}}

@misc{FirstMile,
	author = {{First Mile}},
	howpublished = {online},
	lastchecked = {20 Oct 2025},
	title = {First Mile},
	url = {https://www.thefirstmile.co.uk/},
	year = {2025}}

@article{awino2024,
	author = {Awino, Florence Barbara and Apitz, Sabine E},
	journal = {Integrated Environmental Assessment and Management},
	number = {1},
	pages = {9--35},
	publisher = {Wiley Periodicals, Inc. Hoboken},
	title = {Solid waste management in the context of the waste hierarchy and circular economy frameworks: An international critical review},
	volume = {20},
	year = {2024}}

@inproceedings{bashir2022,
	author = {Bashir, Mohammed JK and Chong, Suk-Ting and Chin, Yun-Tong and Yusoff, Mohd Suffian and Aziz, Hamidi Abdul},
	booktitle = {Solid Waste Engineering and Management},
	howpublished = {[online]},
	pages = {71--164},
	publisher = {Springer},
	title = {Single waste stream processing and material recovery facility (MRF)},
	url = {https://link.springer.com/chapter/10.1007/978-3-030-89336-1_2},
	volume = {2},
	year = {2022}}

@article{chiodo2025,
	author = {Chiodo, Veronica and Gerli, Francesco and Giuliano, Ambra},
	journal = {Journal of Entrepreneurship in Emerging Economies},
	number = {7},
	pages = {156--186},
	publisher = {Emerald Publishing Limited},
	title = {Disentangling tech-enabled system change in social enterprises: an empirical exploration of Ashoka fellows},
	volume = {17},
	year = {2025}}

@techreport{EC2020,
	address = {Brussels},
	author = {{EC}},
	institution = {European Commission (EC)},
	lastchecked = {20 Oct 2025},
	number = {COM(2020)562: 52020DC0562},
	title = {Stepping up Europe's 2030 climate ambition},
	type = {Communication from the Commission to the European Parliament, The Council, The European Economic and Social Coommittee and the Committee of the Regions},
	url = {https://eur-lex.europa.eu/legal-content/EN/TXT/?uri=celex:52020DC0562},
	year = {2020}}

@article{eccles2017,
	author = {Eccles, David W and Arsal, G{\"u}ler},
	journal = {Qualitative Research in Sport, Exercise and Health},
	number = {4},
	pages = {514--531},
	publisher = {Taylor \& Francis},
	title = {The think aloud method: what is it and how do I use it?},
	volume = {9},
	year = {2017}}

@article{edwards2021,
	author = {Edwards, Rosalind and Davidson, Emma and Jamieson, Lynn and Weller, Susie},
	journal = {Quality \& Quantity},
	number = {4},
	pages = {1275--1280},
	publisher = {Springer},
	title = {Theory and the breadth-and-depth method of analysing large amounts of qualitative data: A research note},
	volume = {55},
	year = {2021}}

@article{hahladakis_etal_2018,
	author = {Hahladakis, John N and Velis, Costas A and Weber, Roland and Iacovidou, Eleni and Purnell, Phil},
	journal = {Journal of hazardous materials},
	pages = {179--199},
	publisher = {Elsevier},
	title = {An overview of chemical additives present in plastics: Migration, release, fate and environmental impact during their use, disposal and recycling},
	volume = {344},
	year = {2018}}

@article{hendren2023,
	author = {Hendren, Kathryn and Newcomer, Kathryn and Pandey, Sanjay K and Smith, Margaret and Sumner, Nicole},
	journal = {Public Administration Review},
	number = {3},
	pages = {468--485},
	publisher = {Wiley Online Library},
	title = {How qualitative research methods can be leveraged to strengthen mixed methods research in public policy and public administration?},
	volume = {83},
	year = {2023}}

@article{lee_etal_2022,
	author = {Lee, Donghee N and Hutchens, Myiah J and Krieger, Janice L},
	journal = {Sustainability},
	number = {2},
	pages = {796},
	publisher = {MDPI},
	title = {Resolving the do/do not debate: Communication perspective to enhance sustainable lifestyles},
	volume = {14},
	year = {2022}}

@misc{McBeth_2023_RECOUP,
	address = {Peterborough, UK},
	author = {Tom McBeth},
	howpublished = {online},
	institution = {RECycling of Used Plastics Limited (RECOUP).},
	lastchecked = {10 Jan 2026},
	title = {UK Household Plastic Packaging Collection Survey b2023},
	url = {https://www.recoup.org/wp-content/uploads/2023/12/2023-UK-Household-Plastic-Packaging-Collection-Survey-compressed.pdf},
	urldate = {19 Oct 2025},
	year = {2023}}

@article{nielsen1994_ThinkAloud,
	author = {Nielsen, Jakob},
	journal = {International journal of human-computer studies},
	number = {3},
	pages = {385--397},
	publisher = {Elsevier},
	title = {Estimating the number of subjects needed for a thinking aloud test},
	volume = {41},
	year = {1994}}

@inproceedings{nielsen1994,
	author = {Nielsen, Jakob},
	booktitle = {Proceedings of the SIGCHI conference on Human Factors in Computing Systems},
	pages = {152--158},
	title = {Enhancing the explanatory power of usability heuristics},
	year = {1994}}

@book{osterwalder2015,
	author = {Osterwalder, Alexander and Pigneur, Yves and Bernarda, Gregory and Smith, Alan},
	publisher = {John Wiley \& Sons},
	title = {Value proposition design: How to create products and services customers want},
	year = {2015}}

@article{puvavca2025,
	author = {Puva{\v{c}}a, Nikola and Simin, Mirela Toma{\v{s}} and Bursi{\'c}, Vojislava and {\DJ}uri{\'c}, Katarina and Brki{\'c}, Ivana and Sole{\v{s}}a, Katarina},
	journal = {Journal of Agronomy, Technology and Engineering Management},
	number = {1},
	pages = {1309--1321},
	title = {Economy, European Policies, and Citizens' Behavior: Managing Solid Waste as a Resource},
	volume = {8},
	year = {2025}}

@article{Shooshtarianeetal_2023,
	author = {Shooshtarian, Salman and Maqsood, Tayyab and Wong, Peter S.P. and Caldera, Savindi and Ryley, Tim and Zaman, Atiq and C{\'a}ceres Ruiz, Ana Mar{\'\i}a},
	doi = {10.1108/SASBE-08-2023-0213},
	eprint = {https://www.emerald.com/sasbe/article-pdf/13/2/370/9560104/sasbe-08-2023-0213.pdf},
	issn = {2046-6099},
	journal = {Smart and Sustainable Built Environment},
	number = {2},
	pages = {370-394},
	title = {Circular economy in action: the application of products with recycled content in construction projects - a multiple case study approach},
	volume = {13},
	year = {2023}}

@article{Young_2008,
	address = {New York, NY, USA},
	articleno = {11},
	author = {Young, Indi},
	doi = {10.1145/1376142.1376141},
	issue_date = {April 2008},
	journal = {Ubiquity},
	month = apr,
	numpages = {1},
	publisher = {Association for Computing Machinery},
	title = {Mental Models: Aligning design strategy with human behavior},
	url = {https://doi.org/10.1145/1376142.1376141},
	volume = {2008},
	year = {2008}}

@article{caldwell2008,
	author = {Caldwell, Ben and Cooper, Michael and Reid, Loretta Guarino and Vanderheiden, Gregg and Chisholm, Wendy and Slatin, John and White, Jason},
	journal = {WWW Consortium (W3C)},
	number = {1-34},
	pages = {5--12},
	title = {Web content accessibility guidelines (WCAG) 2.0},
	volume = {290},
	year = {2008}}

@book{krueger2014,
	author = {Krueger, Richard A},
	publisher = {Sage publications},
	title = {Focus groups: A practical guide for applied research},
	year = {2014}}

@book{bowles2010,
	author = {Bowles, Cennydd and Box, James},
	publisher = {Pearson Education},
	title = {Undercover user experience design},
	year = {2010}}

@inproceedings{Bertholdo2016,
	address = {Cham},
	author = {Bertholdo, Ana Paula O. and Kon, Fabio and Gerosa, Marco Aur{\'e}lio},
	booktitle = {Human-Computer Interaction. Theory, Design, Development and Practice},
	editor = {Kurosu, Masaaki},
	isbn = {978-3-319-39510-4},
	pages = {433--444},
	publisher = {Springer International Publishing},
	title = {Agile Usability Patterns for User-Centered Design Final Stages},
	year = {2016}}

@article{bucker2025,
	author = {B{\"u}cker, Christian and Pantel, Julia and Geissdoerfer, Martin and Bhattacharjya, Jyotirmoyee and Kumar, Mukesh},
	journal = {R\&D Management},
	publisher = {Wiley Online Library},
	title = {Digital technologies as enablers of the circular economy: an empirical perspective on the role of companies in driving customer behaviour change},
	year = {2025}}

@article{gibovic2021,
	author = {Gibovic, Denisa and Bikfalvi, Andrea},
	journal = {Recycling},
	number = {2},
	pages = {29},
	publisher = {MDPI},
	title = {Incentives for plastic recycling: How to engage citizens in active collection. Empirical evidence from Spain},
	volume = {6},
	year = {2021}}

@article{pehlken2024,
	author = {Pehlken, Alexandra and Dawel, Lisa and Meyer, Ole and others},
	journal = {Procedia CIRP},
	pages = {563--568},
	publisher = {Elsevier},
	title = {Digital twins: enhancing circular economy through digital tools},
	volume = {122},
	year = {2024}}

@article{dhawan2022,
	author = {Dhawan, Nikhil and Tanvar, Himanshu},
	journal = {Sustainable Materials and Technologies},
	pages = {e00401},
	publisher = {Elsevier},
	title = {A critical review of end-of-life fluorescent lamps recycling for recovery of rare earth values},
	volume = {32},
	year = {2022}}

@article{henry2006,
	author = {Henry, Rotich K and Yongsheng, Zhao and Jun, Dong},
	journal = {Waste management},
	number = {1},
	pages = {92--100},
	publisher = {Elsevier},
	title = {Municipal solid waste management challenges in developing countries--Kenyan case study},
	volume = {26},
	year = {2006}}

@misc{price2020,
	author = {Price, Kaitlin Robb},
	howpublished = {online},
	journal = {WordPress.},
	lastchecked = {20 Oct 2025},
	title = {The millennial wish-cycler: Best practices for reducing recycling contamination},
	urldate = {https://www. jou. ufl. edu/wp-content/uploads/2020/09/The-Millennial-Wish-cycler-Kaitlin-Robb-Price. pdf},
	volume = {[online]},
	year = {2020}}

@article{timlett2008,
	author = {Timlett, Rose E and Williams, Ian D},
	journal = {Resources, Conservation and Recycling},
	number = {4},
	pages = {622--634},
	publisher = {Elsevier},
	title = {Public participation and recycling performance in England: A comparison of tools for behaviour change},
	volume = {52},
	year = {2008}}

@article{konstantinidou2024,
	author = {Konstantinidou, Anna and Ioannou, Konstantinos and Tsantopoulos, Georgios and Arabatzis, Garyfallos},
	journal = {Sustainability},
	number = {22},
	pages = {9969},
	publisher = {MDPI},
	title = {Citizens' attitudes and practices towards waste reduction, separation, and recycling: A systematic review},
	volume = {16},
	year = {2024}}

@article{BONGERS2022,
	author = {Anel{\'\i} Bongers and Pablo Casas},
	doi = {https://doi.org/10.1016/j.ecolecon.2022.107504},
	issn = {0921-8009},
	journal = {Ecological Economics},
	pages = {107504},
	title = {The circular economy and the optimal recycling rate: A macroeconomic approach},
	volume = {199},
	year = {2022}}

@article{anshassi2021,
	author = {Anshassi, Malak and Preuss, Beatriz and Townsend, Timothy G},
	journal = {Journal of the Air \& Waste Management Association},
	number = {8},
	pages = {1039--1052},
	publisher = {Taylor \& Francis},
	title = {Moving beyond recycling: Examining steps for local government to integrate sustainable materials management},
	volume = {71},
	year = {2021}}

@article{mohammed2021,
	author = {Mohammed, Musa and Shafiq, Nasir and Elmansoury, Ali and Al-Mekhlafi, Al-Baraa Abdulrahman and Rached, Ehab Farouk and Zawawi, Noor Amila and Haruna, Abdulrahman and Rafindadi, Aminu Darda'u and Ibrahim, Muhammad Bello},
	journal = {Sustainability},
	number = {19},
	pages = {10660},
	publisher = {MDPI},
	title = {Modeling of 3R (reduce, reuse and recycle) for sustainable construction waste reduction: A partial least squares structural equation modeling (PLS-SEM)},
	volume = {13},
	year = {2021}}

@techreport{EMF2015,
	author = {{EMF} and {McKinsey}},
	institution = {Ellen MacArthur Foundation (EMF) and McKinsey Center for Business and Environment},
	lastchecked = {20 Oct 2025},
	month = {Jan},
	title = {Growth Within: A circular economy Vision for a competitive Europe},
	url = {https://content.ellenmacarthurfoundation.org/m/3d4eba36b0311c08/original/Growth-within-A-circular-economy-vision-for-a-competitive-europe.pdf},
	urldate = {2015},
	year = {2015}}

@techreport{DIT2017,
	author = {{Department for International Development}},
	institution = {Department for International Development},
	month = {March},
	title = {Agenda 2030: The UK Government's approach to delivering the Global Goals for Sustainable Development - at home and around the world},
	url = {https://assets.publishing.service.gov.uk/media/5a75176e40f0b6360e47348f/Agenda-2030-Report4.pdf},
	year = {2017}}

@article{McGrath_BBC2018,
	author = {Matt McGrath},
	journal = {BBC News},
	month = {4 August},
	title = {Plastic food pots and trays are often not recycled, figures show},
	url = {https://www.bbc.co.uk/news/science-environment-45058971},
	volume = {[online]},
	year = {2018}}

@article{vayona2024,
	author = {Vayona, Anastasia and Demetriou, Giorgos and Hartwell, Heather and Britton, Robert and Gillingham, Phillipa},
	journal = {Sustainable development},
	number = {6},
	pages = {6732--6747},
	publisher = {Wiley Online Library},
	title = {A consumer attributions-based approach for investigating the effect of corporate greenwashing on wishcycling},
	volume = {32},
	year = {2024}}

@techreport{DEFRA2024,
	author = {{DEFRA}},
	institution = {Department for Environment Food and Rural Affairs (DEFRA)},
	lastchecked = {20 Oct 2025},
	month = {Nov},
	title = {Simpler Recycling in England: policy update},
	type = {Policy},
	url = {https://www.gov.uk/government/publications/simpler-recycling-in-england-policy-update/simpler-recycling-in-england-policy-update},
	year = {2024}}

@article{rahman2025,
	author = {Rahman, Habib Ur and Khan, Mahiuddin and Ditta, Allah},
	journal = {Bioremediation and Nanotechnology for Climate Change Mitigation},
	pages = {103--136},
	publisher = {Springer},
	title = {Recent Advances in Sustainable Waste Management Practices},
	year = {2025}}

@article{dewasiri2025,
	author = {Dewasiri, Narayanage Jayantha and Singh, Rubee and Kumar, Vikas},
	journal = {Sustainable Innovations and Digital Circular Economy},
	pages = {1},
	publisher = {Springer Nature},
	title = {Decentralised Recycling and Waste Management for Digital Circular},
	year = {2025}}

@article{Oluwadipe2022,
	author = {Saeed Oluwadipe and Hemda Garelick and Simon McCarthy and Diane Purchase},
	doi = {10.1177/0734242X211060619},
	eprint = {https://doi.org/10.1177/0734242X211060619},
	journal = {Waste Management \& Research},
	note = {PMID: 34802336},
	number = {7},
	pages = {905-918},
	title = {A critical review of household recycling barriers in the United Kingdom},
	url = {https://doi.org/10.1177/0734242X211060619},
	volume = {40},
	year = {2022}}

@inproceedings{Mollah2025,
	address = {Singapore},
	author = {MD. Awal Hossain Mollah},
	booktitle = {Service Delivery and the Pulse of Citizen Satisfaction in Urban Governance of Bangladesh},
	doi = {https://doi.org/10.1007/978-981-96-7564-7_5},
	howpublished = {[online]},
	pages = {83 - 101},
	publisher = {Springer Nature},
	title = {Citizen Expectations and Public Service Delivery},
	year = {2025}}

@article{Recycling2004,
	author = {Stephen Tromans},
	journal = {Journal of Enviornmental Law},
	number = {1},
	pages = {80-99},
	title = {Defining Recycling},
	volume = {16},
	year = {2004}}

@misc{WRAP2023,
	author = {{WRAP}},
	howpublished = {online},
	institution = {WRAP},
	lastchecked = {7 Oct 2025},
	month = {March},
	title = {UK Household Food Waste Tracking Survey},
	url = {https://www.wrap.ngo/resources/report/uk-household-food-waste-tracking-survey-autumn-2023},
	year = {2024}}

@techreport{UNSDG,
	author = {{United Nations}},
	institution = {United Nations: Department of Economic and Social Affairs},
	lastchecked = {7 Oct 2025},
	title = {The 17 goals: sustainable development},
	url = {https://sdgs.un.org/goals},
	year = {2015}}

@book{piper2022rubbish,
	author = {Piper, J.},
	publisher = {Unbound},
	title = {The Rubbish Book},
	year = {2022}}

\bibliographystyle{abbrv}

\end{document}